\documentclass[%
 reprint,
 amsmath,amssymb,
 aps,
]{revtex4-1}

\usepackage{pifont}
\usepackage{amssymb,amsmath, amsthm}
\usepackage{mathrsfs,amsmath}
\usepackage{physics}
\usepackage{amsmath}
\usepackage{bbm}

\newtheorem{theorem}{Result}

\usepackage{graphicx}
\usepackage{dcolumn}
\usepackage{bm}

\begin{document}


\title{Majorization in General Grover's Algorithms: Efficient vs Non-efficient cases}

\author{Fernando Mart\' inez Garc\' ia}

\affiliation{%
Departamento de F\' isica Te\' orica I \\ Universidad Complutense. Madrid, Spain.
}%


\begin{abstract}
We study the characterisation of efficient and non-efficient families of Grover's algorithms according to the majorization principle. We develop a geometrical interpretation based on the parameters that appears on these algorithms. Using this interpretation we observe a step-by-step majorization in all efficient Grover's algorithms, whereas the non-efficient Grover's algorithms fail to abide by this majorization principle. These Majorization results are first obtained from numerical calculations. Motivated by these numerical results, we also obtained an analytical demonstration. Finally, the geometrical interpretation found for these parameters is used to obtain additional results of Grover's algorithms that improves our understanding of how they work.
\end{abstract}

\maketitle


\section{\label{sec:level1}Introduction}

Majorization theory emerges as a framework to measure the grade of disorder of classical probability distributions \cite{majorizationbook,libmayoriz}. That one probability distribution majorizes another means that a series of inequalities exist between these probability distributions, i.e, there exists some constraints between these probabilities. From the quantum mechanical point of view, Majorization appears as a solution for a large number of quantum information problems, like conversion of quantum states using local operations \cite{quantstates}.

One of the main open problems in quantum computation theory is finding some mathematical structure underlying optimal quantum algorithms. One possibility for this mathematical structure would be a majorization principle that is satisfied in every optimal quantum computation algorithm. Numerous studies have already been carried out from the majorization point of view, some results are the step-by-step majorization followed in the Quantum Fourier Transformation \cite{QFT} or in complete algorithms, as the one designed to solve the hidden affine function problem \cite{adiab}.

The central piece studied in this work is the well known Grover's algorithm \cite{groverorig} which performs the process of searching in a database of $N$ elements. This is a well-known problem in classical computation for which the best result possible is of $O(N)$ steps \cite{classicsearch}. However, using Grover's algorithm, it is possible to perform this search in $O(\sqrt{N})$ steps.

A simple generalization of Grover's algorithm can be obtained when the database has more than one correct answer for the search. A study of this case has been performed for the case of searching one of $k$ correct elements in a database of $N$ elements \cite{grovermultanswer}. For this case Grover's algorithm requires $O\left( \sqrt{\frac{N}{k}} \right)$ steps. However, we will consider only the case with one correct answer since the generalization for the multiple answer case is straightforward.

Depending  on the values of some parameters, Grover's algorithm can be separated into two types: the efficient ones (which find the desired result with the application of $O(\sqrt{N})$ steps) and the non-efficient ones (which are not capable of finding the desired result) \cite{groverclase}. A study of a particular case of the efficient Grover's algorithm has been performed from the majorization point of view \cite{grover0}, and the conclusion is that this particular algorithm follows a step-by-step majorization.

In this paper we study the most general Grover's algorithm possible, for which a variation of some parameters can be made to obtain an efficient or non-efficient Grover's algorithm. With this in mind, we study these two possible cases to find if they follow a step-by-step majorization process or not. First we study the efficient Grover's algorithms and we obtain that they follow a step-by-step majorization process. Secondly we study the non-efficient Grover's algorithms and we find that they do not follow a step-by-step majorization. Additionally, as a consequence of this study we develop a geometrical interpretation that can be used to understand why efficient Grover's algorithms find the desired result while the non-efficient ones do not.

The main results of this work show that the study of Grover's algorithms supports the existence of a Majorization principle which can be used to distinguish efficient quantum algorithms from the non-efficient ones.

Finally, it is shown that the geometrical interpretation developed in this work can be very helpful in the understanding of how Grover's algorithms behave. To show it we use this interpretation to find already known results but obtained in a different way.

This paper is organised as follows: in Section II we review some basic concepts about majorization theory and its relation to quantum algorithms. We explain the idea of generalized Grover's algorithms and summarize the already known results related to majorization for these algorithms in Section III. In Section IV we develop the mathematical tools we need to perform the majorization study in the most general case. We make a numerical study of generalized Grover's algorithms in Section V, analyzing the effect some parameters of the algorithm have on the majorization behaviour. In Section VI we perform an analytical study only for the efficient algorithm case using an approximation. In Section VII we use the formulas obtained to get some interesting additional results not related with majorization. Finally, we summarize the results obtained and introduce possible ways to continue the study of majorization in quantum algorithms in Section VIII.

\section{\label{sec:level1}Majorization theory in quantum algorithms}

Our approach to the mathematical study of Grover's algorithms will be from the majorization point of view. In this section we explain the relation between majorization and quantum algorithms.

\subsection{\label{sec:level2}Majorization theory}

Let us consider two vectors \textbf{x}, \textbf{y} $\in \mathbb{R}^d$ such that $\sum_{i=1}^d x_i=\sum_{i=1}^d y_i=1$, whose components represent two different probability distributions. Consider the components of these two vectors sorted in decreasing order, denoted as $\textbf{x}^\downarrow$ and $\textbf{y}^\downarrow$. We say that distribution \textbf{y} majorizes distribution \textbf{x}, written as \textbf{x} $\prec$ \textbf{y} if and only if the following relations are satisfied:
\begin{equation}
\sum_{i=1}^k x^\downarrow_i \le \sum_{i=1}^k y^\downarrow_i \qquad k=1,...,d
\end{equation}
probability sums of this kind are called ``cumulants". We must mention that there are other equivalent definitions of majorization \cite{libmayoriz}, but this is the only one needed to understand the results shown in this work.

\subsection{\label{sec:level2}Link between majorization theory and quantum algorithms}

Let $\ket{\psi^{(m)}}$ be the pure state representing the register in a quantum computer at an operating stage labeled by $m=0,...,M-1$, where $M$ is the total number of steps in the algorithm, and let $N$ be the dimension of the Hilbert space. If we denote as $\{\ket{i}\}_{i=1}^N$ the basis in which the final measurement is performed in the algorithm we can associate a set of sorted probabilities $p_i$, $i=1,...,N$, to this quantum state decomposing the register state in the measurement basis, that is:
\begin{equation}
\ket{\psi^{(m)}}=\sum_{i=1}^N a_i^{(m)} \ket{i}
\end{equation}
This way we can associate the following probability distribution to the register state:
\begin{equation}
\textbf{p}^{(m)}=\{p_i^{(m)}\} \quad \text{with} \quad p_i^{(m)} \equiv |a_i^{(m)}|^2
\end{equation}
where $i=1,...,N$. This probability distribution represents the probabilities of all possible outcomes when $m$ steps have been performed. A quantum algorithm will be said to majorize this probability distribution between steps $m$ and $m+1$ if and only if:
\begin{equation}
\textbf{p}^{(m)} \prec \textbf{p}^{(m+1)}
\end{equation}
and from this definition, we can say that a quantum algorithm follows a step-by-step majorization if and only if:
\begin{equation}
\textbf{p}^{(m)} \prec \textbf{p}^{(m+1)} \quad \forall m
\end{equation}
where $m=0,1,...,M-1$.

A step-by-step majorization means that there exists a net flow of probability directed to the values of highest weight, giving way to a steeper distribution with each step of the algorithm. From the physical point of view, this is an evidence of a particular constructive interference which will step-by-step build the solution until reaching the maximum probability for the desired outcome.

It is important to note that the process of majorization is checked on a particular basis, so step-by-step majorization is a basis dependent concept. However, each quantum algorithm is built around a final basis in which the register state will be measured, and each of these possible outcomes have a defined meaning, while other possible basis would not have a meaning for that specific algorithm. This means that there exists a preferred basis in which each algorithm work in order to finally build up an answer, this preferred basis will be the one used to test the step-by-step majorization.

\section{\label{sec:level1}Grover's Algorithm}

In order to study the majorization principle in Grover's algorithm in the most general way, we will need to use a general description of the algorithm. For this purpose, we present a summary of the most important results that appear in the study of Grover's algorithms, the details of the algorithms and their proofs can be found in \cite{groverclase}. After these results are presented, we particularize the results of majorization shown previously for these algorithms.

\subsection{\label{sec:level2}Results of General Grover's Algorithms (GGAs)}

Let us have $n$ qubits, this means that we can create a superposition of $N=2^n$ possible states, the starting point of the algorithm will be a superposition of the $N$ states with equal weight, that is:
\begin{equation}
\label{estadoinicial}
\ket{x_{in}}=\dfrac{1}{\sqrt{N}}\sum_{x=0}^{N-1}\ket{x}
\end{equation}
the aim of the algorithm is to apply operators to this initial state until reaching a maximum probability of finding the result of the search, we will call the sought result $\ket{x_0}$, which is one of the components of the previous summation.

One of the critical results in the study of both Grover's algorithms and their majorization properties insofar as the simplifications it enables, is the fact that the operator from which the Grover's algorithms are formed acts symmetrically on all the states different from $\ket{x_0}$ that compose the summation on (\ref{estadoinicial}). This enables us to work on a system of two dimensions spanned by the following vectors:
\begin{equation}
\begin{aligned}
\ket{x_0} \\
\ket{x_\perp}&\equiv \dfrac{1}{\sqrt{N-1}}\sum^{N-1}_{x\neq x_0}\ket{x}
\end{aligned}
\end{equation}
this way, the initial state can be written as:
\begin{equation}
\ket{x_{in}}=\dfrac{1}{\sqrt{N}}\ket{x_0}+\sqrt{\dfrac{N-1}{N}}\ket{x_\perp}
\end{equation}

In this basis, it is found that the operator used in Grover's algorithms, the so-called Grover kernel, has the following expresion:
\begin{equation}
K=\dfrac{1}{N}\left( \begin{matrix}
1+\delta(1-N) & -\beta(1+\delta)\sqrt{N-1} \\
(1+\delta)\sqrt{N-1} & \beta(1+\delta-N)
\end{matrix}\right)
\end{equation}
were $\beta$ and $\delta$ are complex numbers of modulus 1. Studying the action of $K$ over the initial state one can find \cite{groverclase} that for the Grover's algorithms to be efficient, i.e., to have the chance to achieve a probability near to 1 of finding $\ket{x_0}$, a constraint appear that relates $\beta$ and $\delta$. This constraint is:
\begin{equation}
\beta=\delta\neq -1
\end{equation}
The case that is usually studied is the Grover's algorithm with $\beta=\delta=1$. In this work we study the most general case possible, that is, the so-called General Grover's Algorithms (GGAs) which can have any value of $\beta$ and $\delta$ of modulus 1, including the possibility of having $\beta\neq\delta$ (non-efficient GGAs).

\subsection{\label{sec:level2}Majorization in Generalized Grover's Algorithms}

Now we are in the position of combining the mentioned results of majorization and GGAs. First, the symmetry of the algorithm allows us to work with a simple probability distribution because, as we said before, the symmetry of the problem means that we work in a two dimensional space which at the same time means that the probability distribution is just a vector of two components. This helps because $\textbf{p}^\downarrow$ is then easy to compute: we have one component of the probability distribution, the one associated with $\ket{x_0}$, with a probability $p_{x_0}$ and $N-1$ components, the ones that are not $\ket{x_0}$, with a probability $p_x=\dfrac{1}{N-1}p_{x_\perp}$ where $p_{x_\perp}$ denotes the probability of finding the state $\ket{x_\perp}$, this way it is easy to order the vector \textbf{p} to get $\textbf{p}^\downarrow$.

The fact that there are only two different values of the probability \textbf{p} means that there are only two possibilities for each step, either the state $\ket{x_0}$ transfer part of his probability to the $N-1$ orthogonal states, or all the orthogonal states transfer part of their probability to $\ket{x_0}$. Considering this symmetry of the problem, the only way a majorization process could be satisfied step-by-step until reaching a maximum probability of measuring $\ket{x_0}$ would be if starting from the initial state the probability of measuring $\ket{x_0}$ increases step-by-step.

\subsection{\label{sec:level2}Situation of the majorization study in Grover's algorithms}

Studies of the processes of majorization of Grover's algorithms have already been carried out for the case of $\beta=\delta=1$ \cite{grover0}, and it has been shown that there is, in fact, a step-by-step majorization process satisfied by this particular case of the GGAs.

The aim of this work is to finally settle the question of majorization for all possible GGAs, that is equal to demostrate the existence or not of a step-by-step majorization process for every possible value of $\beta$ and $\delta$ including, of course, the particular case of efficient GGAs (that is, with $\beta=\delta\neq -1$). With the goal of presenting the key variables of GGAs that depend on $\beta$ and $\delta$, we have made a special development of the algorithm that allows us to better understand what changes in the algorithm when we change $\beta$, $\delta$, or both of them.

\section{Key mathematical objects for the study of general Grover's algorithms}

It is known that Grover's algorithm with $\beta=\delta=1$ has a Grover's kernel that works like a rotation in the two dimensional space spanned by $\ket{x_0}$ and $\ket{x_\perp}$ that rotates $\ket{x_{in}}$ to the state $\ket{x_0}$ with sucessive applications of the kernel (that is, through a series of small rotations \cite{rotaciongrover}). The interpretation of Grover's kernel as a rotation means that with each step, the projection of the register state over the vector $\ket{x_0}$ increases which means a step-by-step majorization process.

This is a perfect example of how giving a geometrical meaning to the algorithm simplifies both its understanding and the study of its majorization process. However, the algorithm can no longer be seen as a rotation like this when we have another values of $\beta$ and $\delta$. To get a geometrical meaning to simplify our understanding of GGAs we have made the following development. The register state after $m$ applications of Grover's kernel will be $K^m\ket{x_{in}}$, this means that the probability amplitude corresponding to the state $\ket{x_0}$ after $m$ applications of the kernel, $a_m(x_0)$, is given by:
\begin{equation}
\label{amplitud}
\begin{aligned}
a_m(x_0)\equiv \bra{x_0} K^m \ket{x_{in}}= \\
=\dfrac{1}{\sqrt{N}}\sum_{j=1}^2\{ |\bra{x_0}\ket{\kappa_j}|^2+\sqrt{N-1}&\bra{x_0}\ket{\kappa_j}\bra{\kappa_j}\ket{x_\perp} \}e^{im\omega_j}
\end{aligned}
\end{equation}
where we have introduced the eigenvalues $\xi_{1,2}=e^{i\omega_{1,2}}$ and eigenvectors $\ket{\kappa_{1,2}}$ of the kernel $K$ given by:
\begin{equation}
\xi_{1,2}=e^{i\omega_{1,2}}=\dfrac{1}{2}Tr K \mp \sqrt{-Det K + \dfrac{1}{4}(Tr K)^2}
\end{equation}

\begin{equation}
\label{autovectores}
\ket{\kappa_{1,2}}\propto\left(\begin{matrix}
\dfrac{A\mp\sqrt{-4(DetK) N^2+(TrK)^2N^2}}{2(1+\delta)\sqrt{N-1}} \\ 1
\end{matrix}\right)
\end{equation}
with:
\begin{equation}
A\equiv(\beta-\delta)N+(1-\beta)(1+\delta)
\end{equation}
where the proportionality symbol in (\ref{autovectores}) means that the states are not normalized.

\begin{figure}[ht]
\includegraphics[scale=0.435]{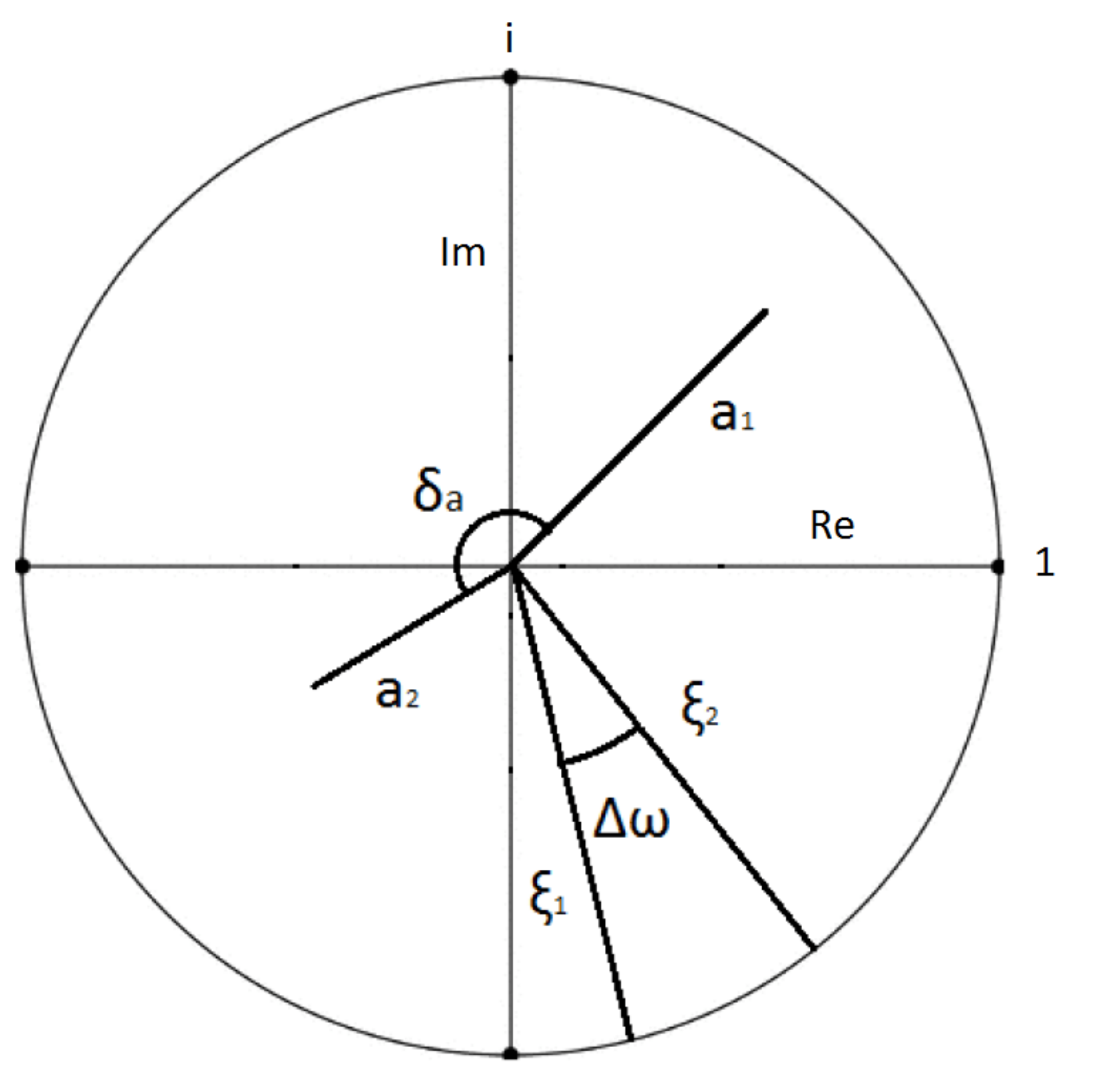}
\centering
\caption{\label{circulo}Example of the representation of the given definitions and their relations. The horizontal axis is the real axis and the vertical is the imaginary axis.}
\end{figure}

However, we are interested in the probabilities, with this in mind we make the following definitions:
\begin{equation}
a_j\equiv\dfrac{1}{\sqrt{N}}(|\bra{x_0}\ket{\kappa_j}|^2+\sqrt{N-1}\bra{x_0}\ket{\kappa_j}\bra{\kappa_j}\ket{x_\perp})
\end{equation}
with $j=1,2$, we will call these as \textit{amplitude components}, and:
\begin{equation}
\delta_a\equiv Arg(a_2)-Arg(a_1)
\end{equation}
which is the \textit{phase difference of the amplitude components}. With these definitions we have that (\ref{amplitud}) can be written as:
\begin{equation}
a_m(x_0)=a_1 e^{im\omega_1}+a_2 e^{im\omega_2}
\end{equation}
and then we have the following expression for the probability of measuring $\ket{x_0}$ at step $m$:
\begin{equation}
p_m(x_0)=||a_1|+|a_2| e^{im(\omega_2-\omega_1)+i\delta_a}|^2
\end{equation}
A representation of these definitions and their relations can be found in the figure \ref{circulo}.

At this point we do not know much about the values of the modulus of $a_1$ and $a_2$ or their phase difference, but there is one thing clear: the probability $p_m(x_0)$ will have a maximum value when $m$ has the value that best aligns $a_1 e^{im\omega_1}$ with $a_2 e^{im\omega_2}$, or equivalently, the value of $m$ which maximizes the real part of $e^{im(\omega_2-\omega_1)+i\delta_a}$. This way it is found that if the sum of the modulus of $a_1$ and $a_2$ takes the value 1 or near to 1 the number of steps $M$ that the algorithm needs to reach the desired result is given by:
\begin{equation}
\label{pasosmaximos}
M=\left[\dfrac{-\delta_a}{\Delta \omega}\right]
\end{equation}
where $\Delta \omega=\omega_2-\omega_1$ and $[q]$ means the natural number closest to $q$.

It is important to notice that both $\delta_a$ and $\Delta \omega$ are only functions of $\beta$, $\delta$ and $N$, i.e., for fixed values of these variables $\delta_a$ and $\Delta \omega$ are constants. If $\delta_a$ is several times greater than $\Delta \omega$ the application of $K$ has the action of rotating $a_1$ and $a_2$ at a constant rate for every application of $K$. This rotation could favour an alignment of these two complex numbers or the opposite, depending on the relative sign of $\delta_a$ and $\Delta \omega$. If these two variables have opposite signs, the kernel will align  better $a_1$ and $a_2$ with each step, which at the same time means an increase of the probability of measuring $\ket{x_0}$ with each step.

This way we have a geometrical interpretation of GGAs which allow us to deduce if there is a step-by-step majorization focusing only on the sign of two variables. The study shown in this work is focused then in the study of the complex numbers $a_1$, $a_2$ and $\Delta \omega$.

\section{Numerical study of the majorization process in Grover's algorithms}

As stated in the previous section, all we need to compute to get an idea of the existence or not of a step-by-step majorization process in a GGA is $\Delta \omega$ and the complex numbers $a_1$ and $a_2$.

Given the different results obtained for the case of efficient and non-efficient GGAs (that is, $\beta=\delta\neq -1$ and $\beta\neq\delta$ respectively) we separate the study in this section into two cases.

\subsection{\label{sec:level2}Efficient algorithms case ($\beta=\delta\neq-1$)}

To make this numerical study, we have parametrized $\beta$ and $\delta$ as:
\begin{equation}
\beta=\delta=e^{it}
\end{equation}

\begin{figure}[ht]
\includegraphics[scale=0.3]{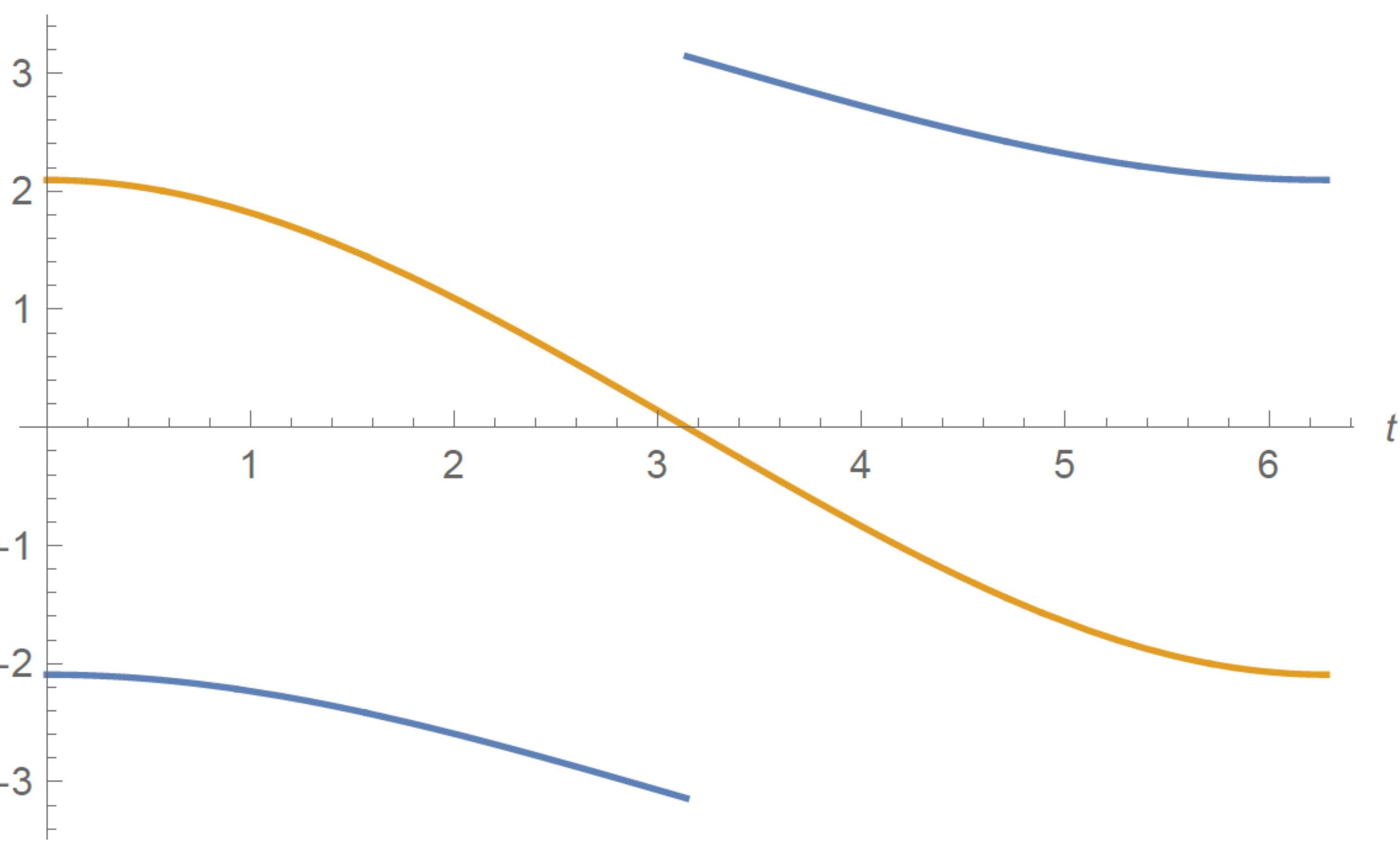}
\centering
\caption{\label{deltaomegaN4eficiente} Representation of the values of $\delta_2-\delta_1$ (in blue) and $\Delta \omega$ (in orange) as functions of the parameter t (from $0$ to $2\pi$) for $N=4$.}
\end{figure}

\begin{figure}[ht]
\includegraphics[scale=0.28]{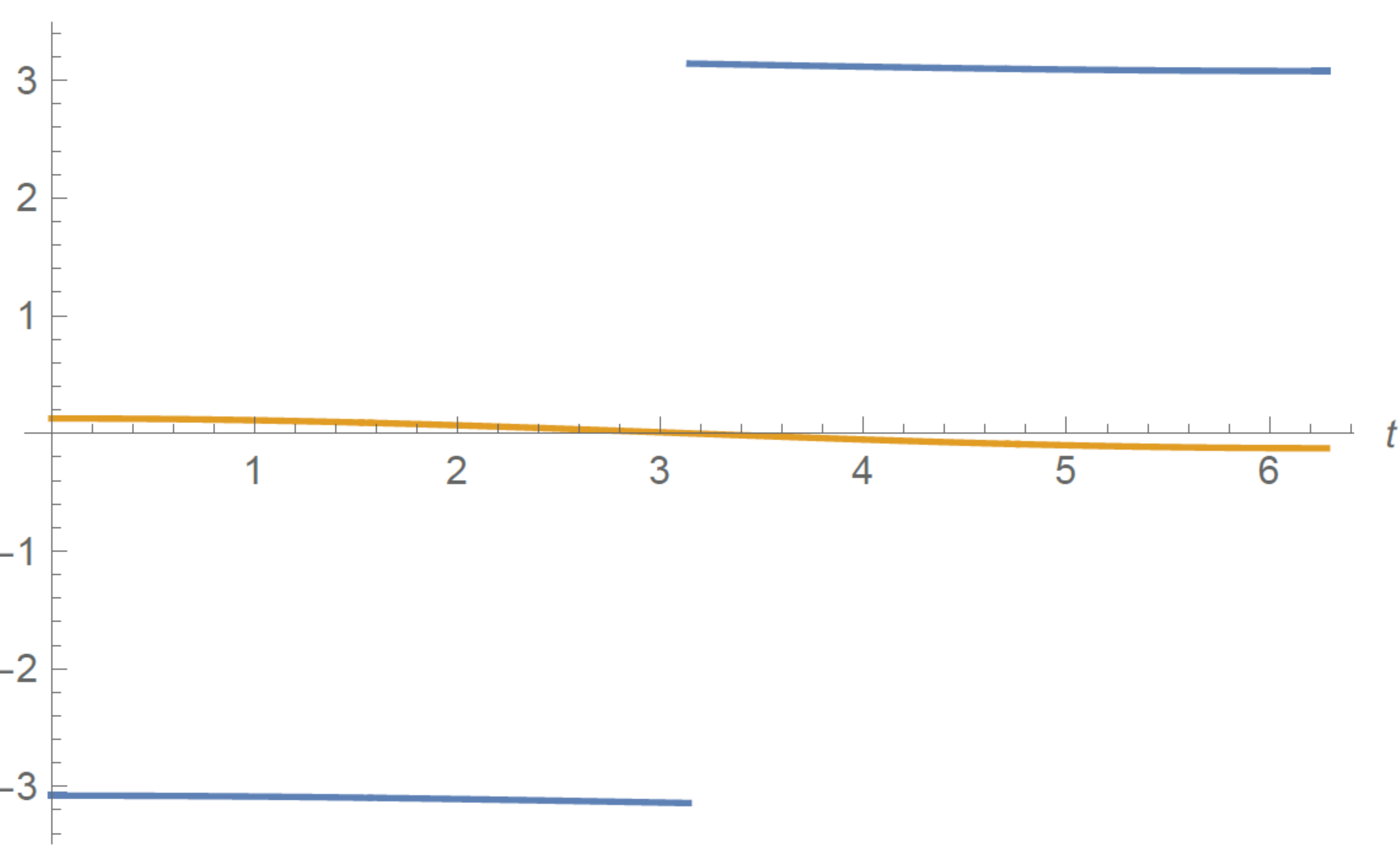}
\centering
\caption{\label{deltaomegaN1000eficiente} Representation of the values of $\delta_2-\delta_1$ (in orange) and $\Delta \omega$ (in blue) as functions of the parameter t (from $0$ to $2\pi$) for $N=1000$.}
\end{figure}

The numerical results obtained for $\delta_a=\delta_2-\delta_1$ and $\Delta \omega$ for all possible $\beta=\delta$ and for $N=4$ and $N=1000$ can be seen in figures \ref{deltaomegaN4eficiente} and  \ref{deltaomegaN1000eficiente} respectively.

It is clear from these figures that the values $\delta_a$ and $\Delta \omega$ have opposites signs for every value of the parameter $t$ (or equivalently, for every value of $\beta=\delta$), and the only effect that varying $N$ has is to change the amplitude of the oscillation-like behaviour these two variables have. It can also be seen that for the case $\beta=\delta=-1$ the value of $\Delta\omega$ is 0, this means that applying the Grover kernel does not produce a rotation. This is related with the fact that for these values of $\beta$ and $\delta$, the Grover kernel is the identity operator.

\begin{figure}[ht]
	\includegraphics[scale=0.60]{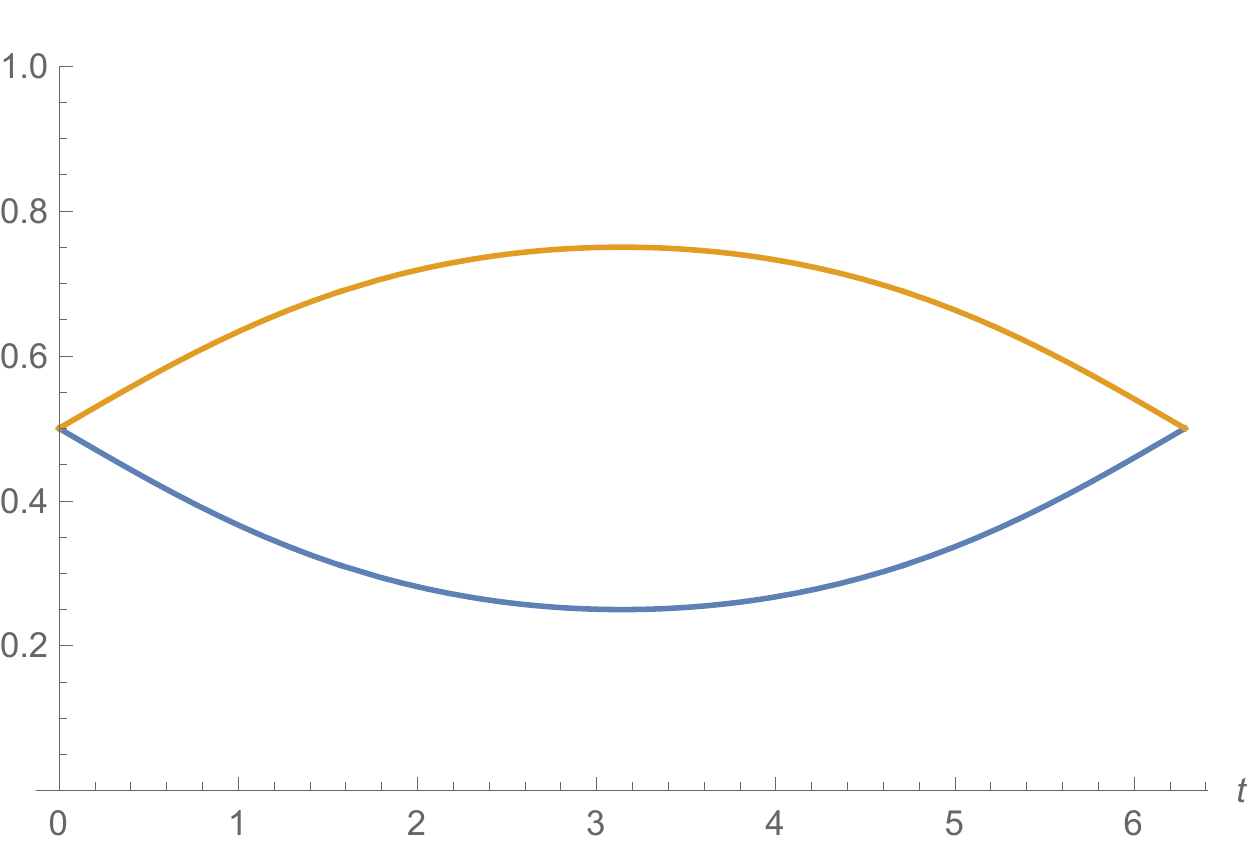}
	\centering
	\caption{\label{a1a2n4} Representation of the values $|a_1|$ (in blue) and $|a_2|$ (in orange) as a function of the parameter t (from $0$ to $2\pi$) for $N=4$.}
\end{figure}

\begin{figure}[ht]
	\includegraphics[scale=0.60]{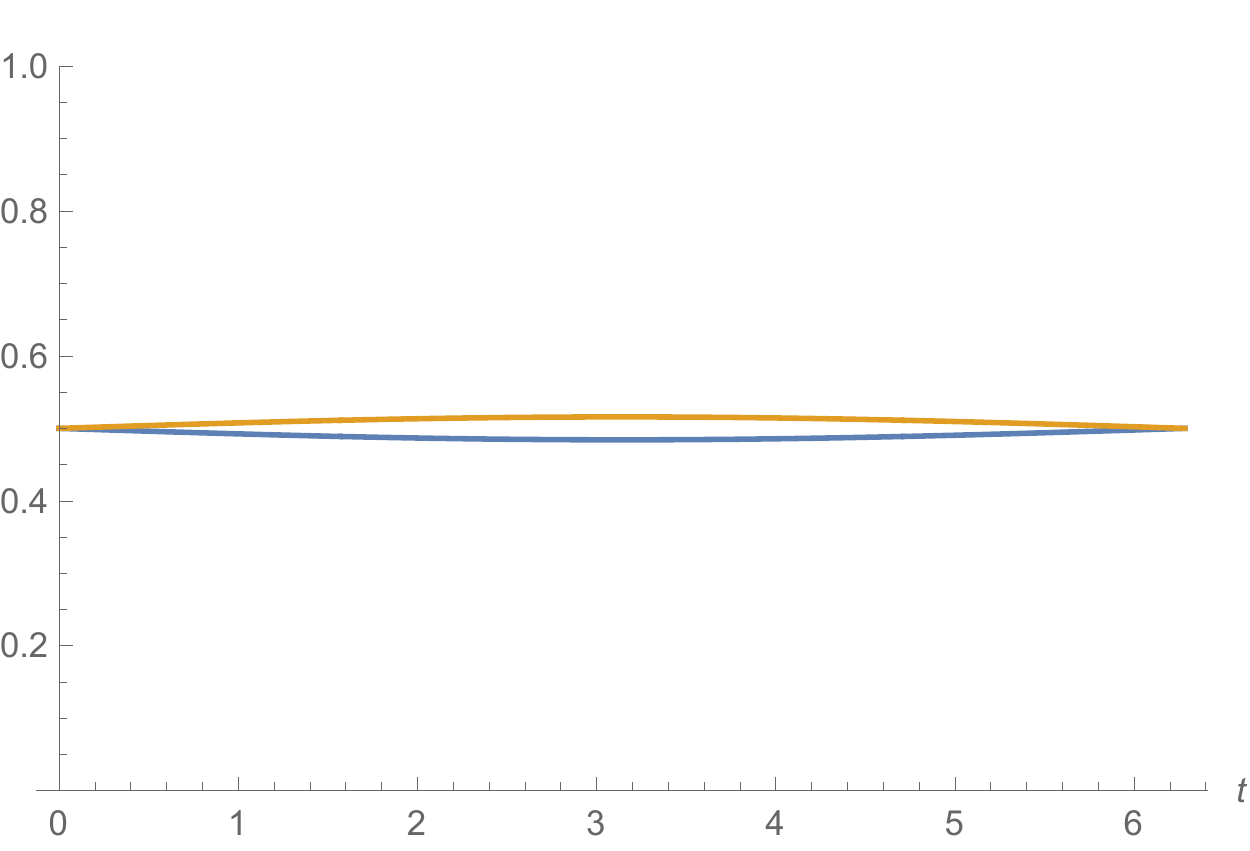}
	\centering
	\caption{\label{a1a2n1000} Representation of the values $|a_1|$ (in blue) and $|a_2|$ (in orange) as a function of the parameter t (from $0$ to $2\pi$) for $N=1000$.}
\end{figure}

We can also see in the figures \ref{a1a2n4} and \ref{a1a2n1000} the variation the modulus of $a_1$ and $a_2$ experiments with the parameter $t$ for $N=4$ and $N=1000$ respectively. One can also notice that in the case we are studying, the values $|a_1|$ and $|a_2|$ have a dependance with $t$ that always satisfies:

\begin{equation*}
|a_1|+|a_2|=1 \quad \forall \beta=\delta
\end{equation*}

This is a very interesting result which shows that every efficient Grover algorithm can get the state $\ket{x_0}$ with a $100\%$ chance if it manages to perfectly align $a_1$ and $a_2$.

We can also derive the known result in which for $N=4$ and $\beta=\delta=1$ only one application of the kernel is neccessary to get the desired state $\ket{x_0}$, this is because for these values $\delta_a$ is exactly $-\Delta \omega$.

Considering these results and the reasoning made in the previous section we can conclude, from a numerical point of view, that all the efficient GGAs perform a step-by-step majorization process because the probability of measuring $\ket{x_0}$ increases step-by-step as a consequence of the opposite sign of $\delta_a$ and $\Delta\omega$.

\begin{figure}[ht]
\includegraphics[scale=0.32]{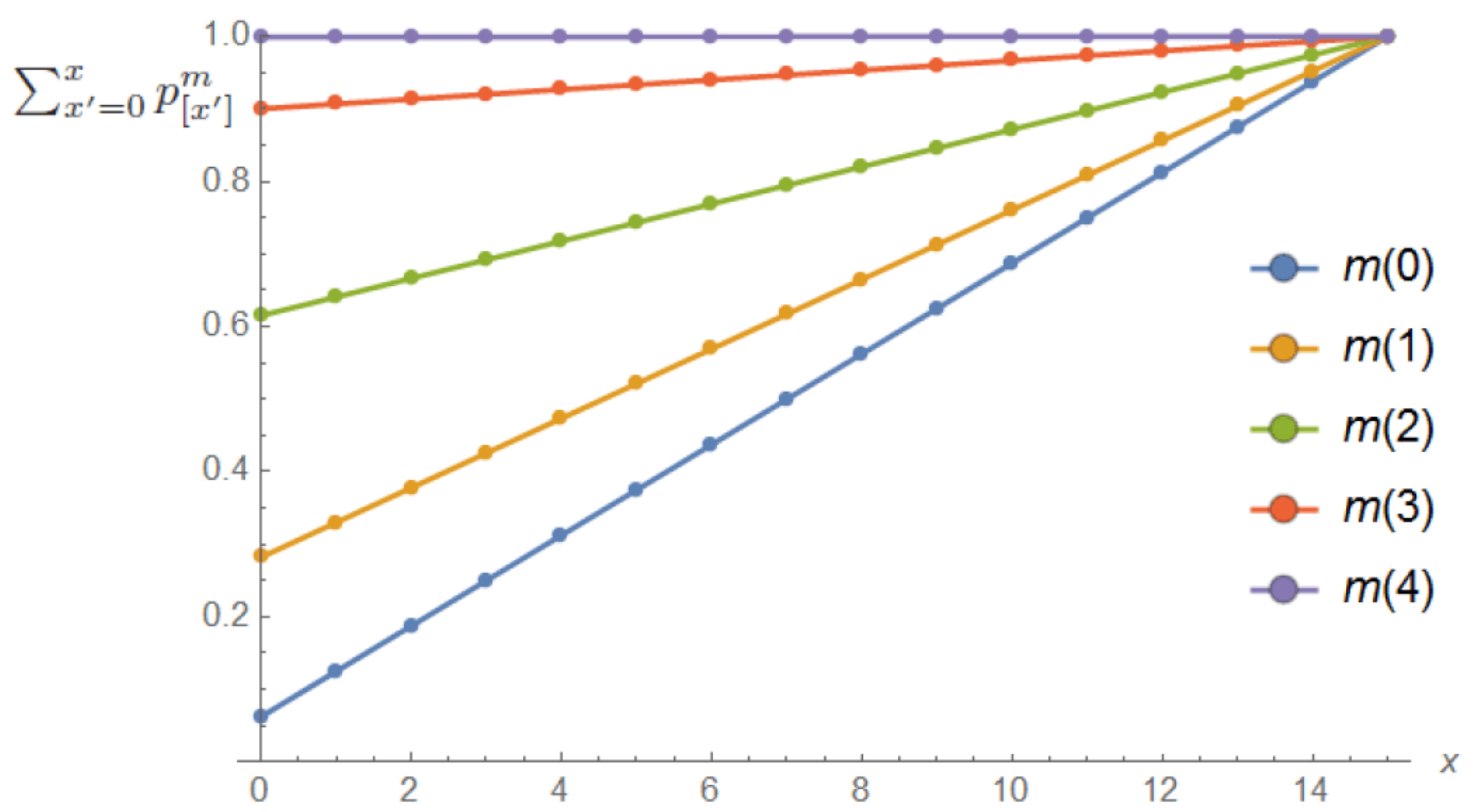}
\centering
\caption{\label{Lorenzsindesfn16tpimedios} Lorenz diagram for $N=16$ and $t=\pi/2$. The values $m(i)$ displayed in the legend identifies the color of each line with the number of applications of the Grover kernel, for example, $m(2)$ identifies the line obtained from the partial summation of the probabilities after two applications of $K$.}
\end{figure}

\begin{figure}[ht]
\includegraphics[scale=0.35]{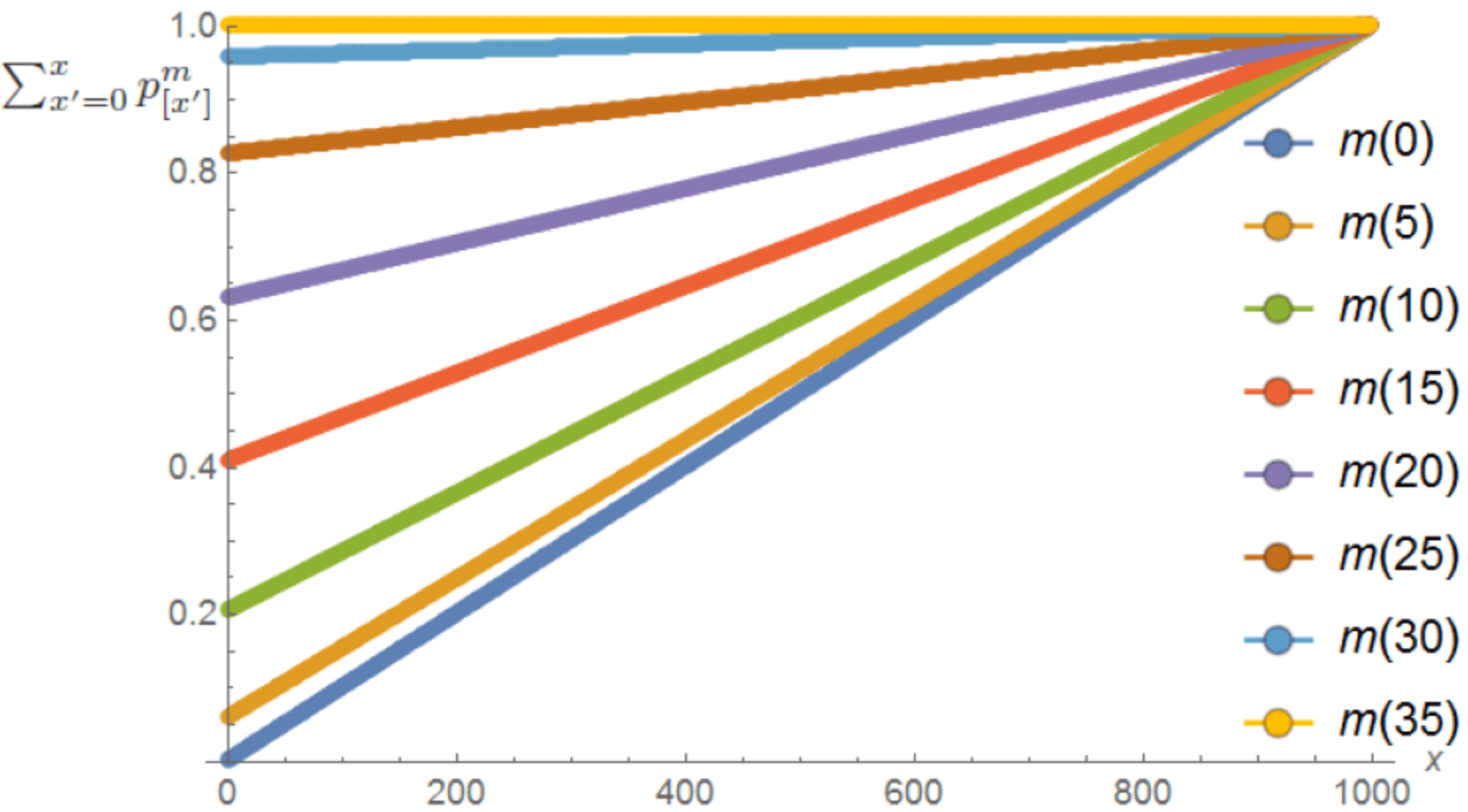}
\centering
\caption{\label{Lorenzsindesfn1000tpimedios} Lorenz diagram for $N=1000$ and $t=\pi/2$. Not all the steps $m$ are displayed to avoid a saturation of the diagram, instead the steps $m$ are shown in intervals of $5$.}
\end{figure}

This process of majorization can be seen directly in figures \ref{Lorenzsindesfn16tpimedios} and \ref{Lorenzsindesfn1000tpimedios} in the so-called \textit{Lorenz diagrams} which show the probability cumulants as a function of the number of elements of $\textbf{p}^\downarrow$, labeled as $x$, considered.

The fact that the functions represented are steeper with increasing values of $m$ and that the lines does not cut each other is a reflection of the majorization process. Thus, we get the following result:

\begin{theorem}
Efficient GGAs (i.e., with $\beta=\delta\neq -1$) always satisfy a step-by-step majorization, independently of the values of $\beta=\delta$ and $N$.
\end{theorem}

\subsection{\label{sec:level2}Non-efficient algorithms case ($\beta\neq\delta$)}

We have already demonstrated numerically that efficient GGAs follow a step-by-step majorization process for all possible values of $\beta=\delta\neq-1$. Now we need to explore the case of non-efficient GGAs, which are given by:
\begin{equation}
\beta\neq\delta
\end{equation}

To perform the study we parametrize the values $\beta$ and $\delta$ as:
\begin{equation}
\begin{aligned}
\beta=e^{it} \\
\delta=e^{ig}
\end{aligned}
\end{equation}
where $t$,$g\in[0,2\pi)$. Of course, extending the study to two independent parameters like this would force us to study the values functions takes over the square $[0,2\pi)\times[0,2\pi)$, i.e., we would have to study a surface. To simplify the study, we will impose a constraint of the kind:
\begin{equation}
\label{consta}
g=t+\text{constant}
\end{equation}
this way, we only have to study a function of one variable, while we still have $\beta\neq\delta$.

\begin{figure}[ht]
\includegraphics[scale=0.6]{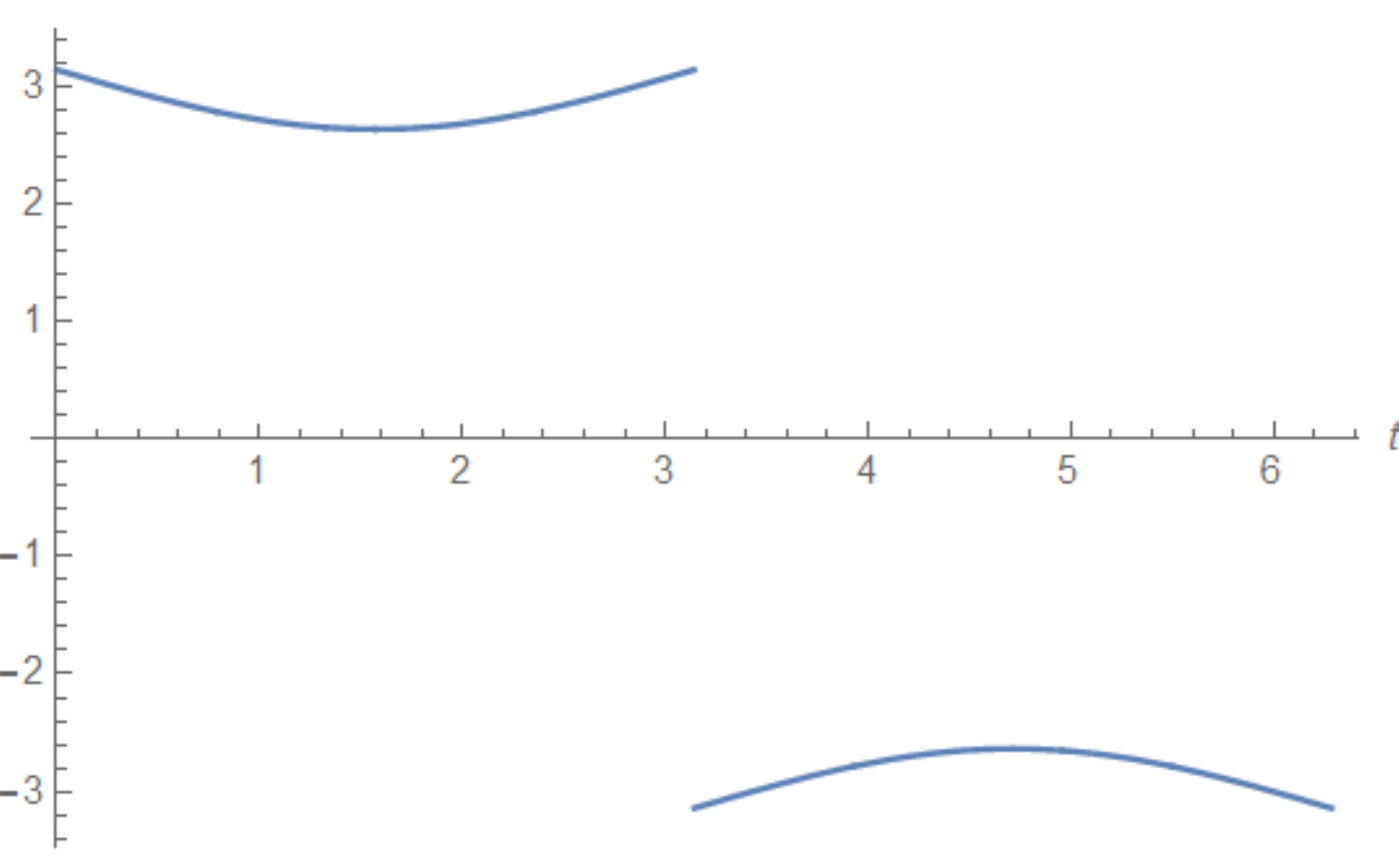}
\centering
\caption{\label{n4desfasepinoeficiente} Representation of $\Delta\omega$ with $N=4$ and $g=t+\pi$ as a function of $t$. The value of $\Delta\omega$ oscillates around the value $\pi$.}
\end{figure}

\begin{figure}[ht]
\includegraphics[scale=0.6]{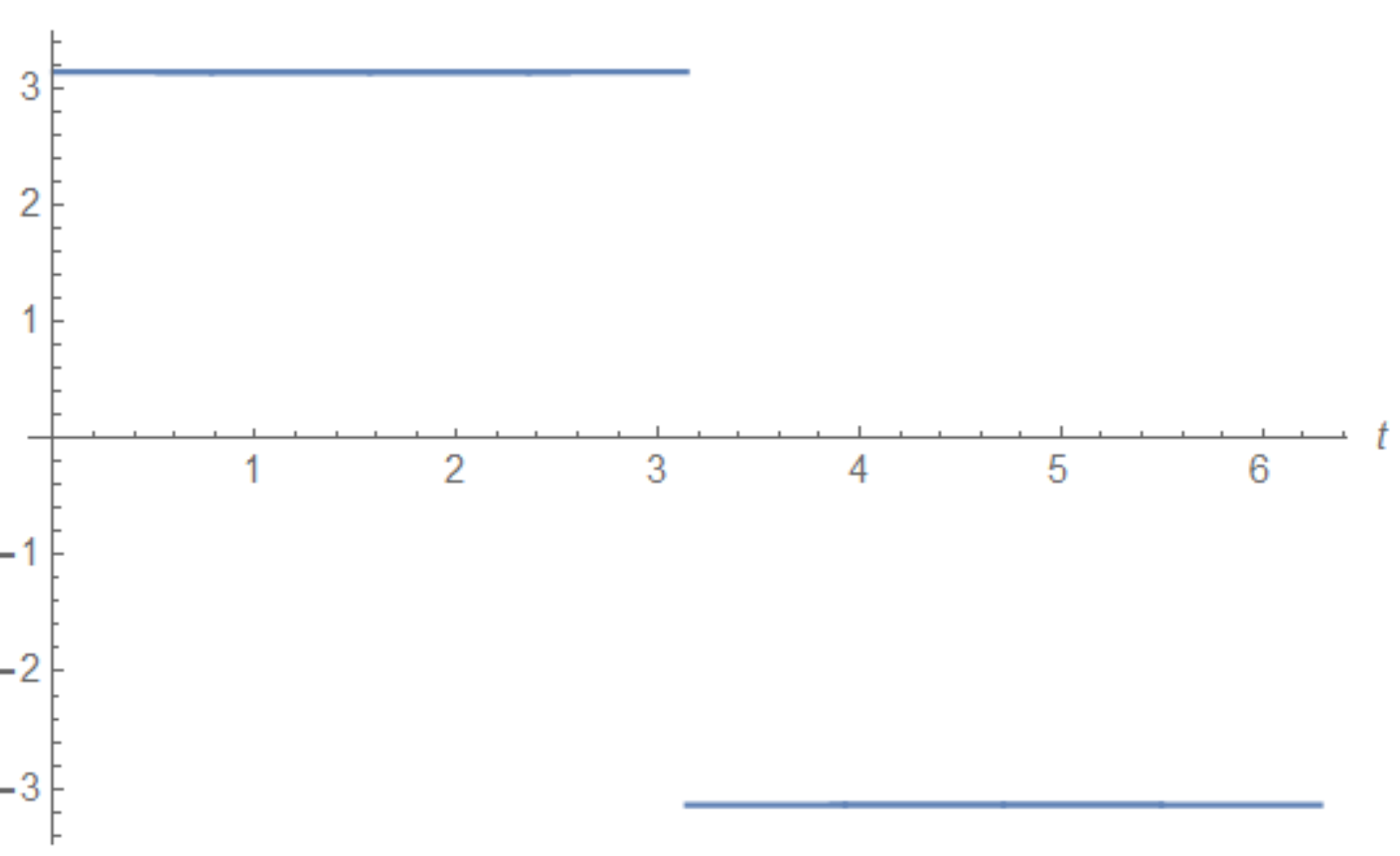}
\centering
\caption{\label{n1000desfpinoeficiente}Representation of $\Delta\omega$ with $N=1000$ and $g=t+\pi$ as a function of $t$. As we can see, the value of $\Delta\omega$ tends to the constant value $\pi$.}
\end{figure}

\begin{figure}[ht]
\includegraphics[scale=0.6]{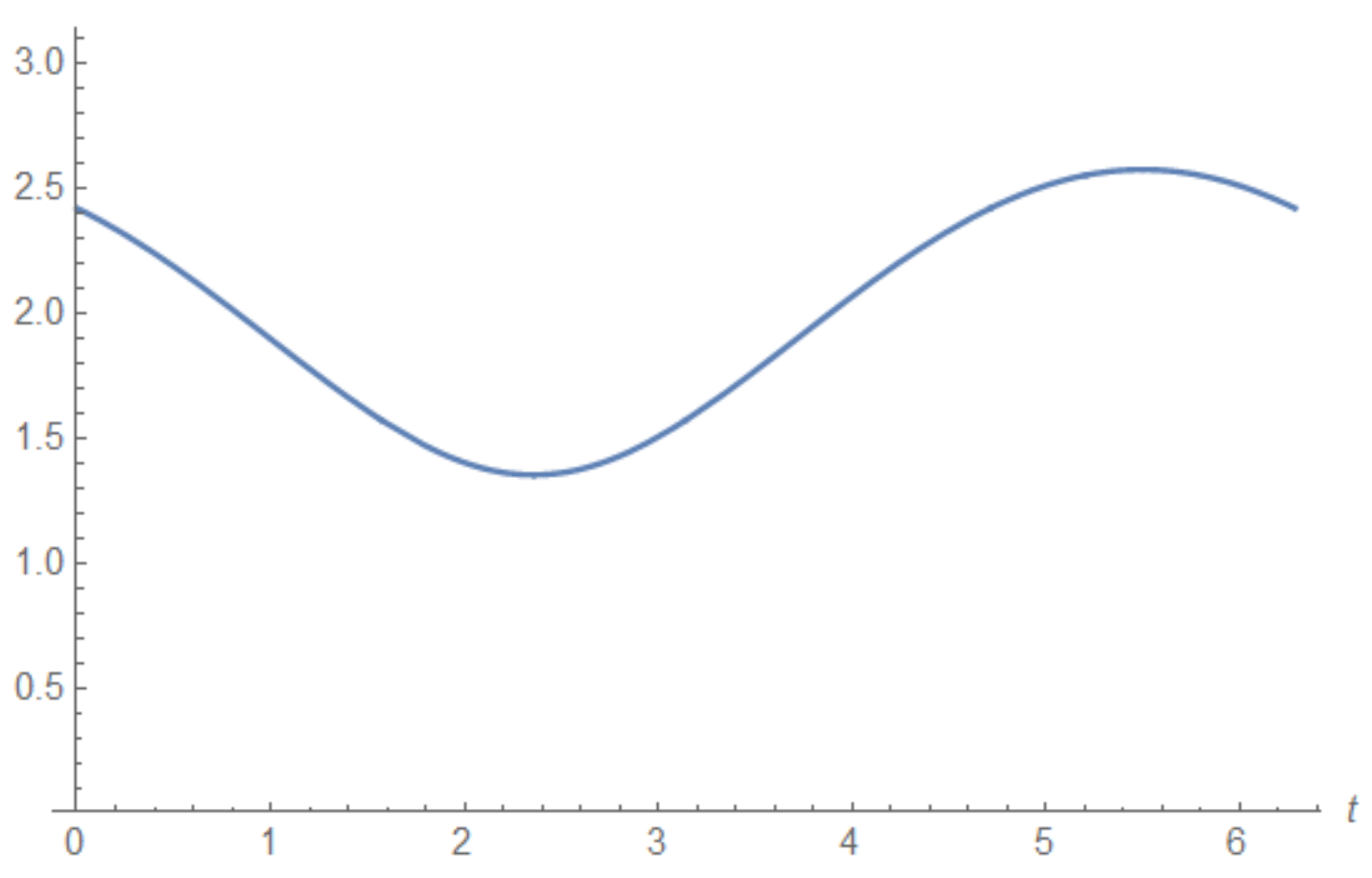}
\centering
\caption{\label{n4desfpimedios} Representation of $\Delta\omega$ with $N=4$ and $g=t+\pi/2$ as a function of $t$. The value of $\Delta\omega$ oscillates around the value $\pi/2$.}
\end{figure}

\begin{figure}[ht]
\includegraphics[scale=0.6]{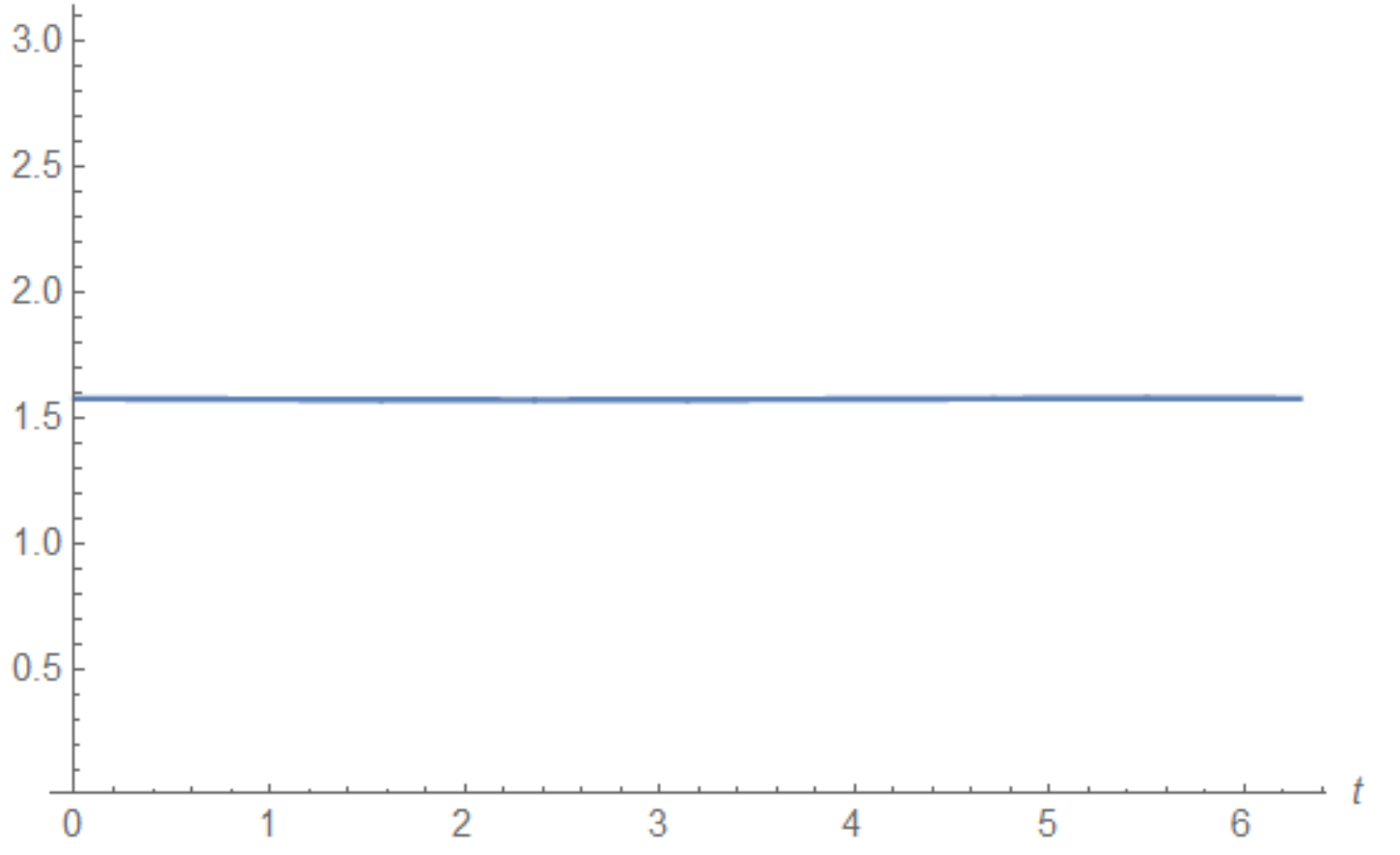}
\centering
\caption{\label{n1000desfpimedios}Representation of $\Delta\omega$ with $N=1000$ and $g=t+\pi/2$ as a function of $t$. As we can see, the value of $\Delta\omega$ tends to the constant value $\pi/2$.}
\end{figure}

We can see the numerical results obtained for $\Delta \omega$ for $g=t+\pi$ and $g=t+\pi/2$ and for $N=4$ and $N=1000$ in figures \ref{n4desfasepinoeficiente} to  \ref{n1000desfpimedios}. Again, the only effect that increasing the value of $N$ has is to reduce the amplitude of the oscillations. Additionally, it can be seen that the values of $\Delta\omega$ tends to the value of the constant introduced in (\ref{consta}) for large values of N.


\begin{figure}[ht]
\includegraphics[scale=0.33]{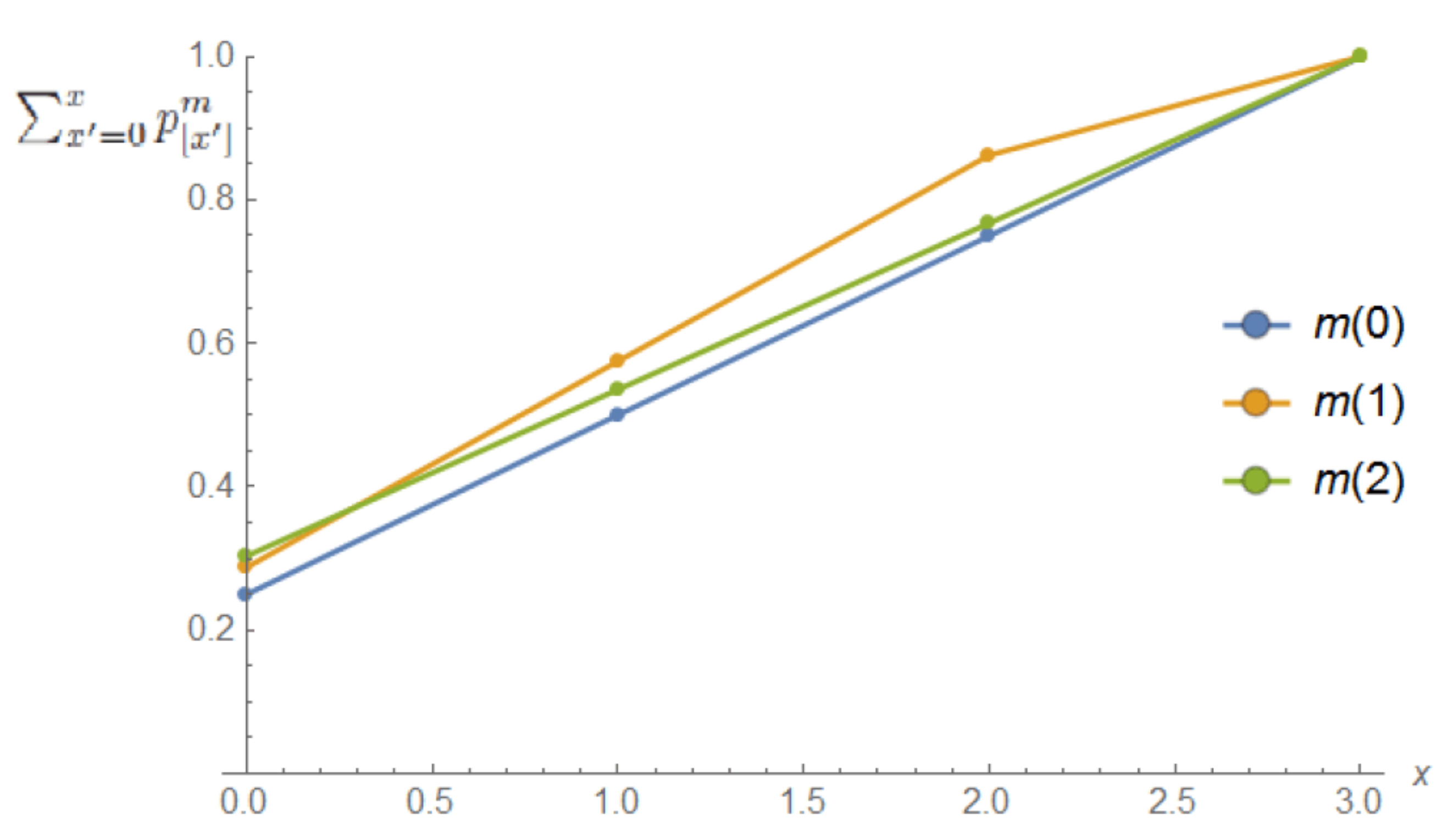}
\centering
\caption{\label{desfn4t1gtpientreunocondos}Lorenz diagram for N=4, $t=1$ and $g=t+\dfrac{5\pi}{6}$. A maximum (at least a relative one) of probability can be seen for $m(2)$. However, the probability distribution of $m(2)$ does not majorize the one of $m(1)$. }
\end{figure}

\begin{figure}[ht]
\includegraphics[scale=0.33]{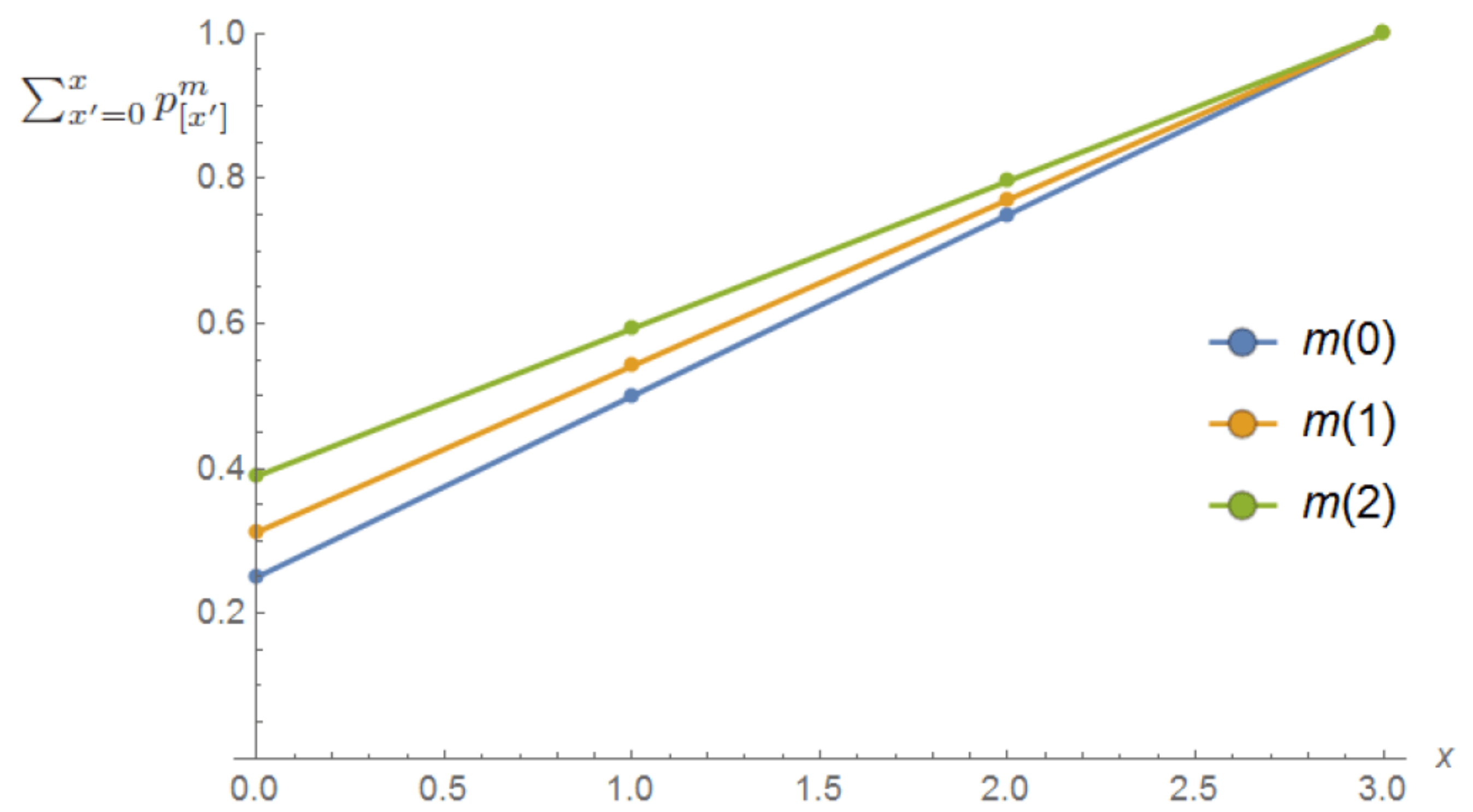}
\centering
\caption{\label{betadeltadif2}Lorenz diagram for N=4, $t=3.37$ and $g=t+\frac{\pi}{3}$. A global maximum of probability is reached at $m(2)$. This diagram might be misleading because it might look like there is a majorization process. This is not true because only a small number of steps is shown, the true chaotic process that takes place when $\beta\neq\delta$ can be seen better in figure \ref{lorenzn100t1gigualtcincosextospi}.}
\end{figure}

Following the steps of the efficient algorithm case, we also show two Lorenz diagrams in figures \ref{desfn4t1gtpientreunocondos} and \ref{betadeltadif2}.


The values of $\delta_a$ are not shown because they are not necessary for the study of majorization given the results of $\Delta\omega$. This is explained as follows: One could wrongly think that  if we restrict ourselves to a part where $\delta_a$ and $\Delta \omega$ have different signs we would get a step-by-step majorization process even with $\beta\neq\delta$. This deduction is wrong, even if the two values have opposite sign, we would not have that $\delta_a$ is several times greater than $\Delta \omega$ (as we can see in the figures, $\Delta\omega$ takes big values, unlike in the efficient algorithm case), and this is vital for the step-by-step majorization process. Having $\delta_a$ and $\Delta \omega$ of opposite signs means that in the first step the probability of measuring $\ket{x_0}$ increases, but if $\delta_a$ is not several times greater than $\Delta \omega$, that means the rotations induced by the kernel would be chaotic, even if the first step increased the probability of measuring $\ket{x_0}$, a second application could apply such a big rotation that it decreased the probability, only to be increased again by a third application and so on.

\begin{figure}[ht]
\includegraphics[scale=0.32]{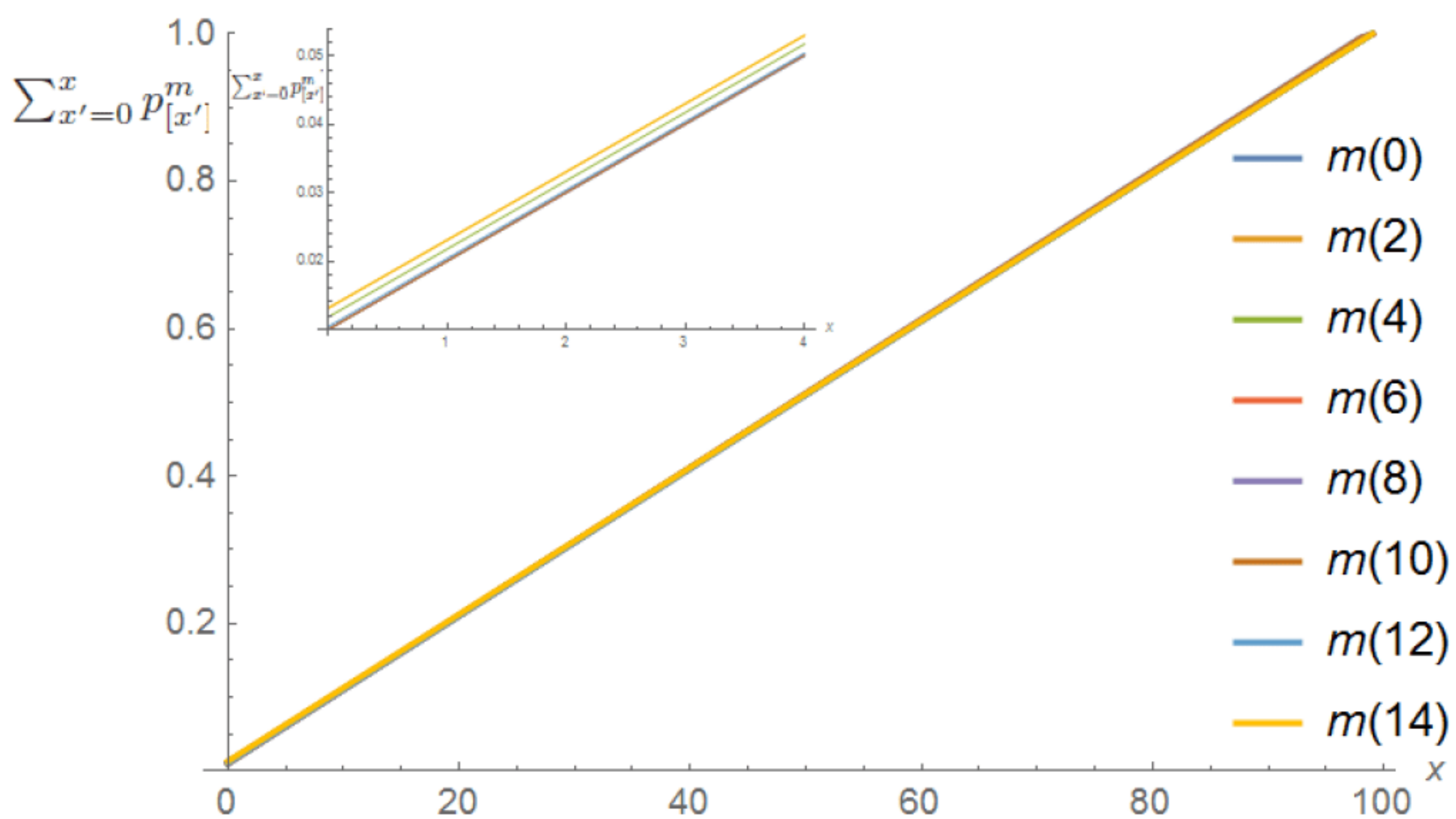}
\centering
\caption{\label{lorenzn100t1gigualtcincosextospi}Lorenz diagram for N=100, $t=1$ y $g=t+\frac{5}{6}\pi$. An inset is shown for the cumulants from $x=0$ to $x=4$. As we can see the algorithm for this case is completely chaotic, so there is no majorization process.}
\end{figure}

In fact, the Lorenz diagrams shown in figures \ref{desfn4t1gtpientreunocondos} and \ref{betadeltadif2} can be misleading because having a small $N$ and a small number of steps $m$ can mask the chaotic process the non-efficient GGAs perform, for which an example is shown in figure \ref{lorenzn100t1gigualtcincosextospi}.

As it can be seen, there is no majorization order in this process, the cause of this beign the big rotation angles that the kernel performs. Thus, we reach the second result of this work:

\begin{theorem}
Non-efficient GGAs (i.e., with $\beta\neq\delta$) do not follow a step-by-step majorization process.
\end{theorem}

\section{\label{sec:level1}Analytical demonstration of majorization for efficient Grover's algorithms}

Motivated by the numerical results shown in the previous section we derive a demonstration of the step-by-step majorization process based on expressions obtained by retaining the terms of order $O\left(\dfrac{1}{\sqrt{N}}\right)$ and ignoring smaller terms. This means that the expressions obtained in this section will be more precise for higher values of $N$.

The approximated expressions for $\Delta \omega$ and $\delta_a$ obtained are the following:
\begin{equation}
\label{deltaomegaaprox}
\Delta\omega\approx 2\arctan\left(\dfrac{2\cos\dfrac{t}{2}}{\sqrt{N}} \right)
\end{equation}
\begin{equation}
\delta_a\approx 2\arctan\left(\dfrac{2\cos\dfrac{t}{2}}{\sqrt{N}} \right)-\pi
\end{equation}

Having in mind that these arctangent have an argument of small value we get:
\begin{equation}
\label{aproxomega}
\Delta\omega\approx \dfrac{4\cos\dfrac{t}{2}}{\sqrt{N}}
\end{equation}
\begin{equation}
\label{aproxdelta}
\delta_a\approx \dfrac{4\cos\dfrac{t}{2}}{\sqrt{N}}-\pi
\end{equation}
which means that when $\Delta\omega$ is positive $\delta_a$ is negative and vice versa. Additionally, $\delta_a$ is several times greater than $\Delta\omega$. This means that the probability of measuring $\ket{x_0}$ increases with each iteration. This way we have analytical arguments which show that there is a step-by-step majorization process in efficient GGAs.

\begin{figure}[ht]
\includegraphics[scale=0.5]{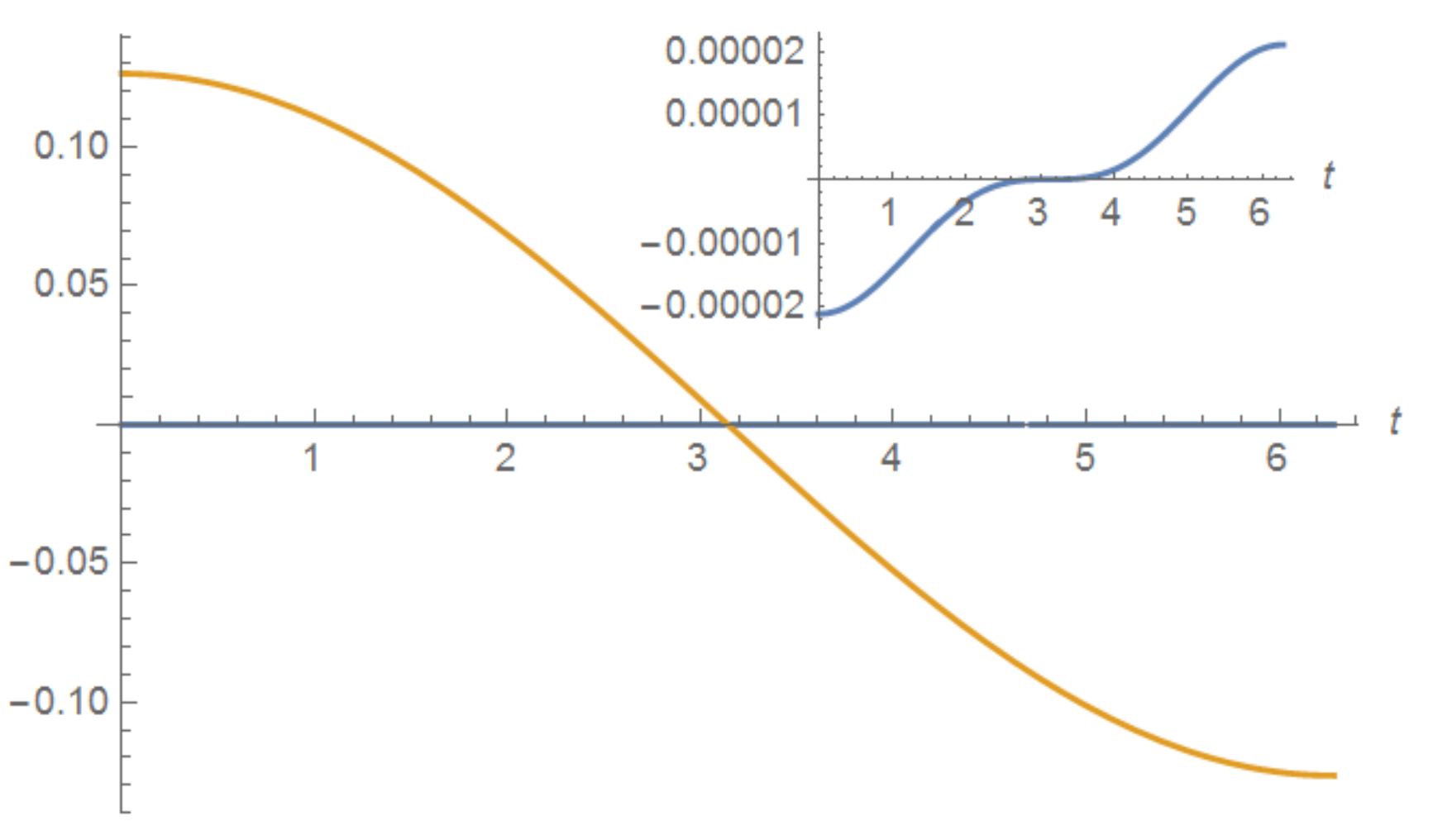}
\centering
\caption{\label{diferroresn1000sindesf}Representation of the approximated formula for $\Delta\omega$ (\ref{deltaomegaaprox}) (in orange) and the difference between this approximated formula and the numerical results for $\Delta\omega$ (in blue) as a function of $t$ for $N=1000$. As it can be seen, the error that appears in the approximation is several orders of magnitude smaller than the value of $\Delta\omega$, so an inset is shown.}
\end{figure}

A representation of the errors caused by taking the approximation is shown in figure \ref{diferroresn1000sindesf} for $N=1000$.
\begin{figure}[ht]
\includegraphics[scale=0.32]{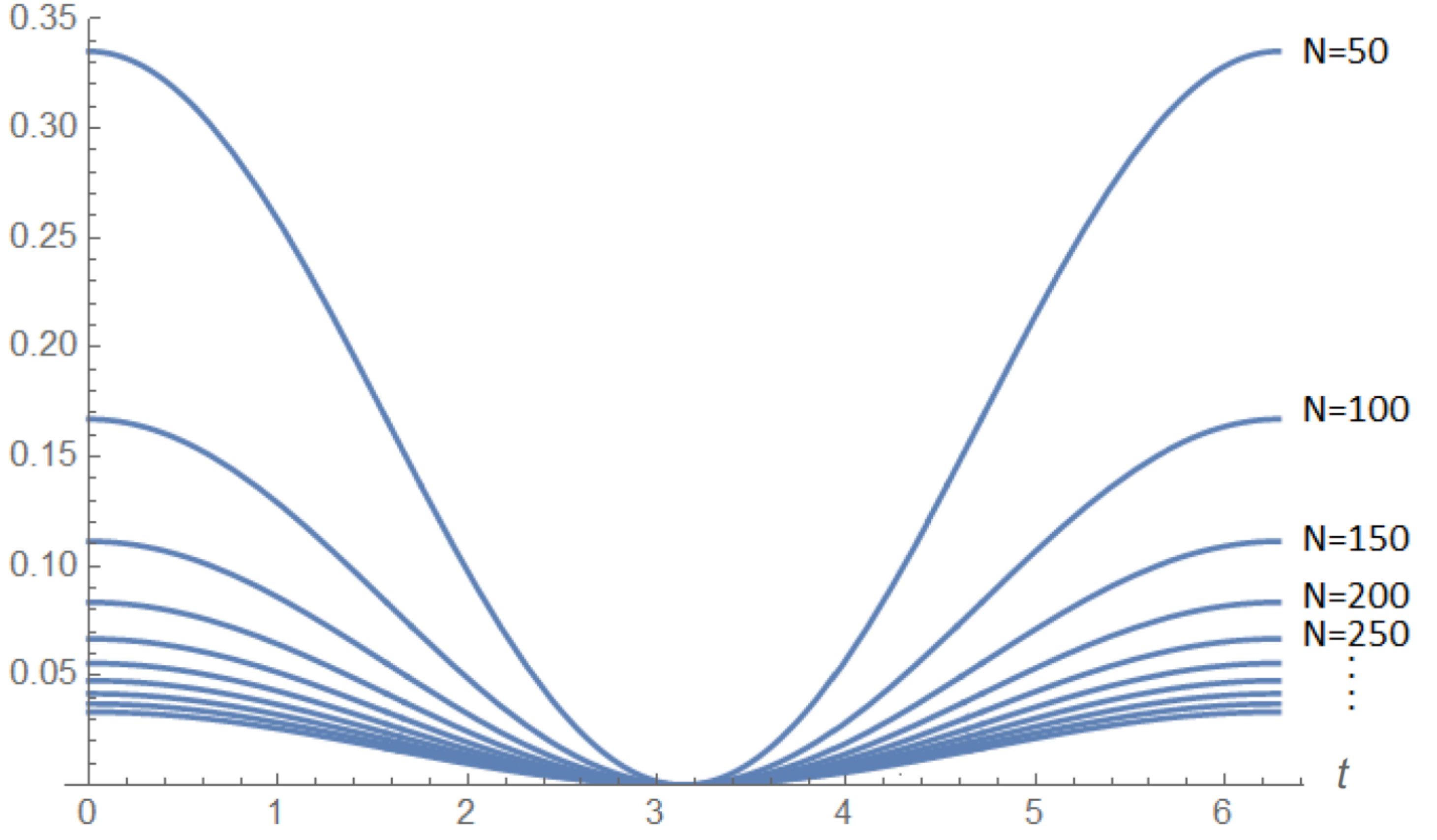}
\centering
\caption{\label{errorrelnpasos50hasta500}Representation of the relative error made with the approximation (\ref{deltaomegaaprox}), i.e., $\dfrac{\Delta\omega_{approx}-\Delta\omega_{numerical}}{\Delta\omega_{numerical}}\cdot 100$ as a function of $t$ for $N$ ranging from $50$ to $500$ in steps of $50$.}
\end{figure}

In figure \ref{errorrelnpasos50hasta500} the relative error of the approximation is shown for several values of $N$. It can be seen that the relative error is reduced for increasing values of $N$, as it should be expected.

\section{\label{sec:level1}Additional results}

The study of the functions $a_1$, $a_2$ and $\Delta\omega$ is not only useful for the study of majorization, it can give insight into the behaviour of GGAs in ways some other mathematical developments of the algorithm can not. As an additional advantage, the study of these functions provides a geometrical meaning of the way the algorithm works, which helps in obtaining an intuitive understanding of the algorithm.

It is shown in \cite{groverclase} that the total number of steps neccessary to complete the algorithm, $M$, for large values of $N$ is given by:
\begin{equation}
\label{pasosclase}
M\approx\left[ \dfrac{\pi}{4\cos\dfrac{t}{2}}\sqrt{N}\right]
\end{equation}
but we have already seen in (\ref{aproxdelta}) that for large values of $N$, $\delta_a$ is approximately $-\pi$, while also having (\ref{aproxomega}). Combining these two results and formula (\ref{pasosmaximos}) we get the same result given by the previous formula, but we give it a clear geometrical meaning: The numbers of steps needed for the algorithm to be complete, $M$, is the number of rotations given by the kernel needed to align $a_2$ and $a_1$.

\begin{figure}[ht]
\includegraphics[scale=0.5]{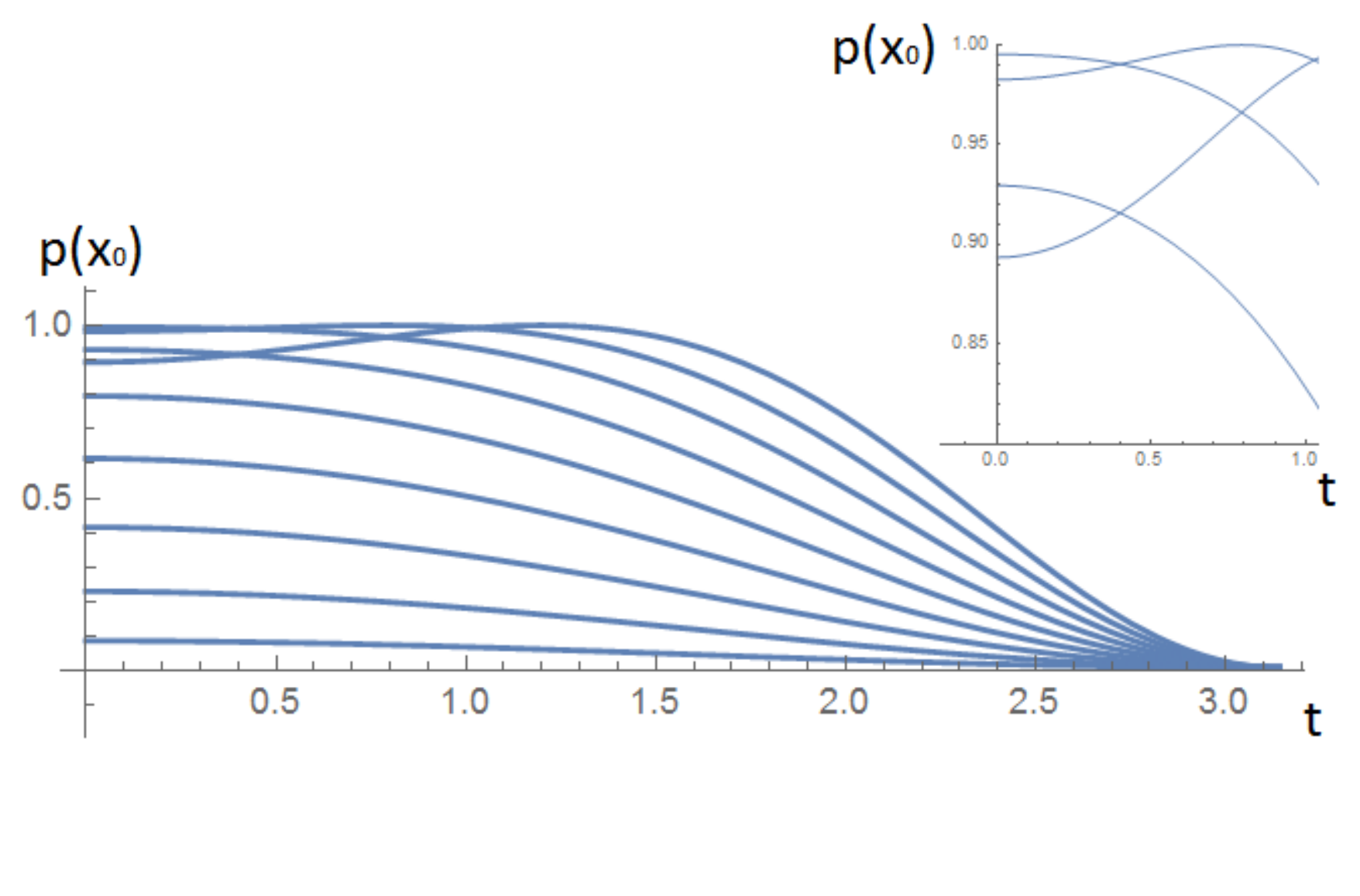}
\centering
\caption{\label{probn100}Representation of the probability of measuring $\ket{x_0}$, $p(x_0)$, as a function of $t$ for $N=100$. The probability increases with each application of the kernel until reaching a maximum.}
\end{figure}

Giving a geometrical meaning to formula ($\ref{pasosclase}$), although it was already a known result, put us on the road to obtain more results. The number of steps $M$ takes the value that best align $a_2$ and $a_1$ but it is possible that even in this case these two values do not end perfectly aligned for a given value of $t$. This situation is shown in figure \ref{probn100}. However, we already know that $\delta_a$ and $\Delta\omega$ are functions of $t$, this means that there exists values of $t$ for which we get:
\begin{equation}
\dfrac{-\delta_a}{\Delta\omega}\in \mathbb{N}
\end{equation}

Thus, varying $t$ we could achieve a situation where the kernel is capable of perfectly align $a_2$ and $a_1$, i.e., the Grover algorithm could achieve a $100\%$ chance of measuring the desired state $\ket{x_0}$ because, as we have seen, in efficient algorithms $|a_1|+|a_2|=1$. This was an already known result, \cite{exactgrover}, but is obtained using a different approach which gives again a geometrical interpretation to this process.

\section{\label{sec:level1}Conclusions and outlook}

Now we can collect all the results found in the analysis of majorization in GGAs:
\begin{itemize}
\item Efficient GGAs follow a step-by-step majorization process due to the ordered behaviour of the rotations applied by Grover's kernel in these cases.
\item Non-efficient GGAs do not follow a step-by-step majorization process due to the chaotic rotations applied by their Grover's kernel.
\end{itemize}
These results support the idea of a Majorization principle as an underlying mathematical structure working behind the efficient algorithms. It is important to note that the step-by-step majorization only occurs in the efficient case until the maximum probability is achieved. After this point, the rotations applied by Grover's kernel disalligns $a_1$ and $a_2$, which leads to a decrease in the probability of measuring the desired state step-by-step. This means that the majorization process only occurs until the algorithm has find the desired result.

Additionally, we have shown that the mathematical development obtained in this work can be of great help in the understanding of GGAs. In fact, this geometrical interpretation of the GGAs can be used to obtain an algorithm that works with a 100\% chance of success. In the same way, this geometrical interpretation is really interesting to develop an intuition of how the algorithm works.

Finally, it is worth mentioning that this work can be continued with a study of the algorithm performed by real devices. This way, if a Majorization principle exists, majorization could be used to detect which experimental implementation of the algorithm performs a truly  efficient process. Some studies of a majorization principle in quantum computation in experimental devices has been carried out with very promising results \cite{exp1}.

\end{document}